\documentstyle[aps,prb,preprint]{revtex}

\begin{document}
\draft
\title{S and D Wave Mixing in High $T_c$ Superconductors}
\author{C. O'Donovan and J.P. Carbotte}
\address{Department of Physics and Astronomy,
McMaster University, Hamilton, Ontario, Canada L8S 4M1 \\
preprint: arch-ive/9502035 \\
email: odonovan@mcmaster.ca}
\date{February 1, 1995}
\maketitle

\begin{abstract}
For a tight binding model with nearest neighbour attraction
and a small orthorhombic distortion, we find a phase diagram for the
gap at zero temperature which includes three distinct regions as a
function of filling.  In the first, the gap is a mixture of mainly
$d$-wave with a smaller extended $s$-wave part.  This is followed by a
region in which there is a rapid increase in the $s$-wave part
accompanied by a rapid increase in relative phase between $s$ and $d$
from 0 to $\pi$.  Finally, there is a region of dominant $s$ with a
mixture of $d$ and zero phase.  In the mixed region with a finite
phase, the $s$-wave part of the gap can show a sudden increase with
decreasing temperature accompanied with a rapid increase in phase
which shows many of the characteristics measured in the angular
resolved photoemission experiments of Ma {\em et al.} in
$\rm Bi_2Sr_2CaCu_2O_8$
\end{abstract}


\narrowtext


Many experiments in the last few years have been interpreted in terms
of a gap with $d_{x^2-y^2}$ symmetry in the $\rm CuO_2$ planes.  Among
these is the penetration depth experiment of Hardy et al.\cite{hardy} which
show a linear dependence over a large temperature range down to the
lowest temperatures measured.  This is illustrative of a large class
of experiments indicating that the gap goes to zero on the Fermi
surface but which are not sensitive to its phase.  Many new
experiments have been devised specifically to test the
phase\cite{wollman,brawner,tsuei,sun} of
the gap over the Brillouin zone with mixed results.  Some favour
$d_{x^2-y^2}$ symmetry, others extended $s$-wave.  Perhaps the most
detailed information on the momentum dependence of the gap has come
from a variety of angular resolved photoemission
data.\cite{olson,shen,hwu,kelley,desseau,ding1} Recently,
Ding et al.\cite{ding2} have found that the zero's in the gap are displaced
somewhat from the main diagonals to either side of the $(\pi,\pi)$
direction.  This can arise if the gap is not pure $d_{x^2-y^2}$ but
has an additional small $s$-wave component with zero relative phase
between the two.  This can be achieved if a small orthorhombic
distortion is introduced in the $\rm CuO_2$ plane\cite{odonovan,kazuhiro} and
in
addition the bilayer nature\cite{liu,jian} of $\rm Bi_2Sr_2CaCu_2O_8$ is taken
into account to give nodes on both sides of the diagonal.  In the
experiment of Ma et al.,\cite{jian} the temperature variation of the gap is
investigated in different directions in the Brillouin zone.  In
particular, the gap is found small and perhaps zero near $T_c$ in the
$(\pi,\pi)$ direction and to increase rapidly below a reduced
temperature of $T/T_c = 0.8$.  In the $(0,\pi)$ direction, the gap is
large and follows more closely a {\sc bcs} behaviour.

     In this paper, we investigate the phase diagram for simple {\sc
bcs} model with first nearest neighbour attraction with a view at
understanding its phase diagram as a function of filling $n$ and of
making connections with the experimental situation described above.
For the band structure, we employ a tight binding model with first and
second nearest neighbour hopping.  For tetragonal symmetry, we find at
zero temperature, as a function of decreasing filling a region of pure
$d_{x^2-y^2}$ symmetry which is followed by one of mixed $d_{x^2-y^2}$
plus an extended $s_{x^2+y^2}$ part with relative phase of $\pi/2$
between the two pieces.  Finally at yet lower filling, there is a
region of pure $s_{x^2+y^2}$ gap.  If some small but finite
orthorhombic distortion is introduced into the $\rm CuO_2$ plane, the
situation changes considerably in the sense that no pure $d_{x^2-y^2}$
region remains.  Instead, it is replaced by a mixture of mainly
$d_{x^2-y^2}$ but with a small admixture of $s_{x^2+y^2}$ with no
relative phase so that in this region, the nodes in the gap are
shifted somewhat off the $(\pi,\pi)$ direction.  As the filling is
lowered at some particular critical value of $n$, the relative phase
starts growing sharply from 0.  In this regime there are only point
nodes in the gap at $(\pm\pi/2,\pm\pi/2)$ in the Brillouin zone of the
${\rm CuO_2}$ plane. This increase in relative phase is accompanied by
a rapid increase in the $s_{x^2+y^2}$ part and similar sharp decrease
in $d_{x^2-y^2}$.  When the phase has increased to $\pi$, the $s$-wave
part dominates and the gap will again have zeros on the Fermi
surface.  Adding an on site repulsion to the model changes the range
of the various regions somewhat but more distinctively adds a small
constant component to the gap which can be of the same or opposite
sign as the $s_{x^2+y^2}$ part.



     The {\sc bcs} gap equation determines the energy gap or order
parameter $\Delta_{\bf k}$ as a function of momentum ${\bf k}$ in the
reciprocal lattice.  To describe electrons in a 2-dimensional $\rm
CuO_2$ plane, we will use a tight binding dispersion relation with up
to second nearest neighbours of the form

\begin{eqnarray}
\varepsilon_{\bf k}&=&-2t\left[ \cos(k_x a)+(1+\delta)\cos(k_y
a)\right.\nonumber \\
&-&\left. 2B\cos(k_x a)\cos(k_y a)-(2-2B-\mu) \right]
\label{dispersion.eq}
\end{eqnarray}

\noindent with $t$ the first nearest neighbour hopping parameter which is often
assumed to be $100 \rm meV$, $B$ second neighbour hopping in units of
$t$ often taken to be $B=0.45$ for $\rm Y_1Ba_2Cu_3O_7$ and $\mu$ is
the chemical potential.  The final parameter $\delta$ in equation (1)
is an orthorhombic distortion introduced into the $\rm CuO_2$ plane
which could be thought of as roughly modeling the existence of chains
in $\rm YBaCuO$.  It has almost become conventional to eliminate the
chemical potential in Eq.~\ref{dispersion.eq} in favour of the filling
factor $n$ defined as

\begin{equation}
 n=\frac{2}{\cal N}\sum_{\bf k}{\left[1 - {\varepsilon_{\bf k} \over E_{\bf k}}
\tanh\left({1 \over 2} \beta E_{\bf k}\right)\right]}
 \label{filling.eq},
\end{equation}

\noindent where the sum over $\bf k$ is over the first Brillouin zone of the
$\rm CuO_2$
plane and ${\cal N}$ is a normalization factor so that half filling
corresponds to $n=1$.

In momentum space, the {\sc bcs} gap equation is given in terms of the
pairing potential $V_{\bf k-k^\prime}$ and the temperature
$\beta^{-1}=k_B T$ by

\begin{equation}
\Delta_{\bf k} = \frac{1}{\cal N}\sum_{\bf k^\prime}{\left[V_{\bf k -
k^\prime}{\Delta_{\bf k^\prime} \over 2 E_{\bf k^\prime}} \tanh\left(
\frac{1}{2} \beta E_{\bf k^\prime}\right)\right]},
\label{bcs.eq}
\end{equation}

\noindent with $E=\sqrt{\varepsilon_{\bf k}^2+\Delta_{\bf k}^2}$ the
quasiparticle energies in the superconducting state.

The nearest neighbour part of the pairing potential has the form:

\begin{equation}
V_{\bf k-k^\prime}^{(nn)}=2g_x\cos(k_x-k_x^\prime)+2g_y\cos(k_y-k_y^\prime),
\label{interaction.eq}
\end{equation}

\noindent where $a_x$ and $a_y$, the lattice parameters, are both taken to
be $a$. In Eqs.~\ref{dispersion.eq} and \ref{interaction.eq} it can be
set equal to one if the momentum is measured in units of $2\pi/a$.
In Eq.~\ref{interaction.eq}, $g_x$ and $g_y$ are coupling constant
taken to be equal and their value set by the desired value of critical
temperature $T_c$ or of gap amplitude.  With the form in
Eq.~\ref{interaction.eq} for the pairing potential, the {\sc bcs}
equation, Eq.~\ref{bcs.eq}, can be solved by convolution using fast
Fourier transform techniques using

\begin{equation}
 \Delta_{\bf k}=\frac{1}{\cal N}\sum_{\bf k^\prime}{V_{\bf k - k^\prime}F_{\bf
k^\prime} }
={\cal F}^{-1}\left[ {\cal F}\left[ V_{\bf k^\prime}\right] \cdot {\cal
F}\left[ F_{\bf k^\prime}\right] \right],
\end{equation}

\noindent where ${\cal F}[x_{\bf k}]\equiv \sum_{\bf k}e^{\imath {\bf
k}\cdot{\bf r}}x_{\bf k}$ is the Fourier transform operator and $F_{\bf
k}\equiv\Delta_{\bf k}\tanh\left(\beta E_{\bf k}/2\right)/2E_{\bf k}$.

Another technique for solving the gap equation, Eq.~\ref{bcs.eq}, with the
pairing potential of Eq.~\ref{interaction.eq} is to expand the
interaction in terms of separable functions:

\begin{eqnarray}
V_{\bf
k-k^\prime}^{(nn)}&=&2g_x\{\cos(k_x)\cos(k_x^\prime)-\sin(k_x)\sin(k_x^\prime)\} \nonumber \\
&+&2g_y\{\cos(k_y)\cos(k_y^\prime)-\sin(k_y)\sin(k_y^\prime)\}. \nonumber \\
\label{separate.eq}
\end{eqnarray}

The $\sin(k_{x,y})$ terms do not contribute for singlet
pairing.  For a general separable interaction of the form:

\begin{eqnarray}
V_{\bf k - k^\prime}^{(sep.)}=\sum_{n,m} v_{n,m}\eta_{\bf k}^n \eta_{\bf
k^\prime}^m
\label{basis.eq}
\end{eqnarray}

\noindent of which Eq.~\ref{separate.eq} is a special case, we can write
Eq.~\ref{interaction.eq} as:

\begin{equation}
\Delta_{\bf k}=\sum_{n}\eta_{\bf k}^n \sum_{m} v_{n,m} \frac{1}{\cal
N}\sum_{\bf k^\prime}{\eta_{\bf k^\prime}^m{F_{\bf k^\prime}}},
\label{bcs.sep.eq}
\end{equation}

\noindent and we see that $\Delta_{\bf k}$ has no more
components $\eta_{\bf k}^n$ than there are in the interaction.  We
will assume the $\eta_{\bf k}^n$ to be an orthonormal set and only a
few are needed to represent Eq.~\ref{bcs.sep.eq}.  The fast Fourier
transform technique for solving Eq.~\ref{bcs.eq} serves as a check on
the alternate method represented in Eq.~\ref{bcs.sep.eq} which
involves a set of coupled integral equations which can be solved by
vector integration.  Before presenting results (similar work can be
found in refernece \cite{spathis,micnas,dahm}) for the solution of the gap
equation, we present,
for completeness, the expression for the penetration depth
$\lambda_{i,j}^{-2}(T)$ which is a tensor.  Results for this quantity will
be given in the next section.  The expression for $\lambda_{i,j}^{-2}(T)$
is:

\begin{equation}
 \lambda_{ij}^{-2}\propto\sum_{\bf k}\frac{\partial \varepsilon_{\bf
k}}{\partial k_i}\frac{\partial \varepsilon_{\bf k}}{\partial k_j}\left[
\frac{\partial f(E_{\bf k})}{\partial E_{\bf k}}-\frac{\partial
f(\varepsilon_{\bf k})}{\partial \varepsilon_{\bf
 k}}\right]
 \label{pd.eqn}
\end{equation}

\noindent where the $\frac{\partial \varepsilon_{\bf k}}{\partial k_i}$
are related to Fermi velocities and the overall constant in
Eq.~\ref{pd.eqn} is of no interest here.


In Fig.~\ref{DvsN.fig}, we show a phase diagram for the gap at zero temperature
as
a function of filling $n$ (see Eq.~\ref{filling.eq}) for a case $t=100
\rm meV$, $B=0.45$, $T=0$ and $g_{x,y}=0.75t$.  A small orthorhombic
distortion $\delta=0.1$ is also included in the calculations.
Throughout the phase diagram, only the two components
$d_{x^2-y^2}\propto (\cos(k_x)-\cos(k_y))$ and $s_{x^2+y^2} \propto
(\cos(k_x)+\cos(k_y))$ appear in the gap.  The amplitude multiplying
these two functions are denoted by solid circles ($d_{x^2-y^2}$ part)
and solid square ($s_{x^2+y^2}$ part).  The relative phase $\phi$
between the two basis functions is also given.  We note that in the
region down to approximately $n=0.45$, the solution at zero
temperature $T=0$ is mainly $d$-wave with a small admixture of
extended $s$-wave.  The relative phase between these two components is
zero throughout this region which means that the zero line nodes of
the gap will be displaced somewhat off the main diagonals.  This is
seen in the angular resolved photoemission data of Ding et al.(13)
although our model includes a single plane and nodes are found only on
one side of the diagonals and not on both sides as seen in the
experiments.  This can be understood, however, quite naturally if a
bilayer model were used \cite{kazuhiro,liu} to model $\rm Bi_2Sr_2CaCu_2O_8$
instead of the single plane model used here.  The small admixture of
$s_{x^2+y^2}$ just described is due entirely to our use of a small
orthorhombic distortion in the electronic dispersion \cite{hardy} and the
amplitude of this component would be zero in a pure tetragonal case so
that we would get instead, in this case, a region of pure $d_{x^2-y^2}$ in our
phase
diagram.

Next, we note in Fig.~\ref{DvsN.fig} that below approximately $n=0.45$ up to
about
$0.28$, we have a region of admixture of extended s-wave and
$d_{x^2-y^2}$ coexistence but with a finite relative phase of $\phi$.  For
$\phi=\pi/2$, the admixture is roughly of equal proportion while a
small relative phase corresponds to a small amount of extended $s$ and
a phase near $\pi$ a small amount of $d_{x^2-y^2}$.  Below $n\approx
0.28$, we recover a region of zero relative phase but with the two
amplitudes having opposite signs and the $s_{x^2+y^2}$ part dominating.
If an orthorhombic distortion was not included in our calculation the
phase would always be $\pi/2$ but the $s_{x^2+y^2}$ component would
only set in at some critical hole concentration $n_{c1}$ and the
$d_{x^2-y^2}$ component would drop to zero sharply at some other smaller
critical concentration $n_{c2}$ with $n_{c1}$ and $n_{c2}$ dependent
on the parameters of the theory such as details of the pairing
potential and the underlying band structure.

     The region of finite phase (not zero or $\pi$) in Fig.~\ref{DvsN.fig} is
very
interesting and was investigated further.  For a fixed value of
filling $n$ taken to be $0.36$, we show in Fig.~\ref{DvsT.fig} our results for
the
two components of the gap as a function of temperature.  At high
temperatures near the critical temperature $T_c$, we have a sharply
rising $d_{x^2-y^2}$ (solid circles) component as $T$ is reduced and a
small $s_{x^2+y^2}$ (solid squares) component.  There is zero phase
between these two components.  But abruptly at $T \approx 36\rm K$ to
be compared with a critical temperature of approximately $68\rm K$ the
phase switches on and the $s_{x^2+y^2}$ component starts rising quite
rapidly.  Our results are in qualitative agreement on these points
with the angular resolved photoemission data of Ma et al.\cite{jian} Of
course our phase diagrams only roughly approximate the observed doping
dependance. In view of this we make no attempt to fit the data since
our model is quite simplified and, in particular, does not include the
bilayer nature of the $\rm CuO_2$ planes in $\rm Bi_2Sr_2CaCu_2O_8$,
and so in the high temperature regime, our gap nodes fall only on one
side of the diagonals.  At lower temperature, once there is a finite
relative phase between the two components, the line node in the gap at
the Fermi surface is lost, of course, because it is no longer real and
both real and imaginary components would need to vanish
simultaneously.  For a $d_{x^2-y^2}$ and $s_{x^2+y^2}$ mixture, this
can happen only at four points $(\pm\pi/2,\pm\pi/2)$ which, in
general, will not fall on the Fermi surface.  This can be seen in much
more detail in Fig.~\ref{DvsK.fig}, but before doing so, we point out the
insert in
Fig.~\ref{DvsT.fig} in which we show the unnormalized penetration depth in the
two
principles in plane directions as a function of temperature.  Since
$\delta\neq 0$ in our calculations, $\lambda_{xx}^{-2}$ and
$\lambda_{yy}^{-2}$ will be different although, as can be seen from
the figure, their temperature dependence tracks each other fairly well
and both show a clear change in slope at $T=36\rm K$, the point at
which the relative phase $\phi$ becomes finite.

In Fig.~\ref{DvsK.fig}, we show contour plots of the gap in the two dimensional
$\rm CuO_2$ plane Brillouin zone as well as a plot of the Fermi
surface.  Four frames are shown.  Two correspond to the region above
the second critical temperature $T_{c2} \approx 36{\rm K}$ and two
below.  Above $T_{c2}$, the phase between the two components of the
gap is zero and so the order parameter is real and its value can be
plotted as a function of $k_x$ and $k_y$.  We see in Fig.~\ref{DvsK.fig}
(frame b) positive and negative contours for $\Delta(T=40{\rm K})$ and
that the zero contour is not a straight line along the diagonal as it
would be for pure $d_{x^2-y^2}$ wave.  Instead, it is displaced below
the diagonal around $(\pi,\pi)$ and above it around $(0,0)$. Only
at $(\pi/2,\pi/2)$ is it exactly on the diagonal.  Referring again to
Fig.~\ref{DvsT.fig}, it is clear that the amount the node in the gap at the
Fermi surface will
have moved off the main diagonals will depend on temperature.  In
frame (d) of Fig.~\ref{DvsK.fig}, we show $|\Delta(T=40{\rm K})|$ rather than
its
actual value.  This graph differs from frame (b) only in as much as
all contours are for positive values of the gap.  It is to be used for
easier comparison with the data of frame (a) which is for
$|\Delta(T=30{\rm K})|$.  In this case, the gap is complex and there
is no zero contour for $|\Delta(T=30{\rm K})|$.  The amount of change
in the contours as compared to the previous frame (d) is not large,
however, since the temperature has not been changed by much.  The final
set of contours shown in frame (c) are for low temperature and now
these are quite different.  We note, in particular, that
$|\Delta(T=1{\rm K})|$ has a value near $6 \rm meV$ at the Fermi
surface in the direction $(\pi,\pi)$ and about $16 \rm meV$ for the
$(0,\pi)$ direction in rough agreement with the experimental situation
found in $\rm Bi_2Sr_2CaCu_2O_8$.(17)


Using a simple tight binding model for the electron dispersion
relation and a pairing potential involving a nearest neighbour
attraction, we have calculated a phase diagram for the gap as a
function of filling including in the calculation a small orthorhombic
distortion.  At high filling, the gap is found to be mainly of
$d_{x^2-y^2}$ symmetry with a small admixture of extended s- wave
($s_{x^2+y^2}$) less than 10\% in our work.  The phase between the two
components at zero temperature is zero so that in this region of phase
space there are line nodes and the zeros of the gap in the Brillouin zone are
slightly displaced off the diagonal.  The degree of displacement
increases with decreasing filling.  In our work, the $s_{x^2+y^2}$
component is completely due to the orthorhombic distortion and
disappears in a tetragonal system.  At some critical filling $n_{c1}$,
a phase develops between $d_{x^2-y^2}$ and $s_{x^2+y^2}$ components.
As the phase grows rapidly with decreasing filling, the amount of $d_{x^2-y^2}$
component
decreases rapidly and the amount of $s_{x^2+y^2}$ increases.  At some
lower critical filling $n_{c2}$ not so different from $n_{c1}$, the
phase becomes $\pi$ and remains at this value for $n \leq n_{c2}$ in
which case the state is mainly $s_{x^2+y^2}$ with a smaller admixture
of $d_{x^2-y^2}$.  In a tetragonal system, the concentrations $n_{c1}$
and $n_{c2}$ would be between the boundary of pure $d_{x^2-y^2}$ and
mixture $d+\imath s$ and between $d+\imath s$ region and pure $s$ wave
respectively with the phase in the mixed region exactly $\pi/2$.

In the region of finite phase $\phi$ for the zero temperature gap, the
temperature variation of the gap is complicated.  At high temperature
near $T_c$ (equal to $68{\rm K}$ for the case considered), the gap is
real and is an admixture of mainly $d_{x^2-y^2}$ with a small amount
of $s_{x^2+y^2}$ so that the zero gap contours are displaced from the
main diagonals.  At some critical temperature equal to $36\rm K$ in
our example, a finite phase develops between the s and d components.
While the $d_{x^2-y^2}$ component changes little at this temperature
and shows an overall temperature dependence that is similar to that of
an isotropic superconductor, the $s_{x^2+y^2}$ abruptly increases with
decreasing temperature as observed in the angular resolved
photoemission experiments of Ma et al.(17) In this temperature region,
the gap shows no zeros on the Fermi surface and in our model ends up
at $T=0$ with $s_{x^2+y^2}$ part roughly half the $d_{x^2-y^2}$ part.
These ratios depend on the parameters used of course.  Calculation of
the temperature dependent penetration depth show a clear change in
slope of both the $x$ and $y$ components of the in plane penetration
depth, $\lambda_{xx}^{-2}$ and $\lambda_{yy}^{-2}$, at $T=36{\rm K}$.


Research supported in part by the Natural Sciences and Engineering
Research Council of Canada ({\sc nserc}) and by the Canadian Institute
for Advanced Research ({\sc ciar}).  We thank P. Soininen for
discussions and interest.

\begin{figure}
\caption{Phase diagram for the gap at zero temperature as a function of
hole concentration with $n=1$ corresponding to half filling.  The
parameters used in the calculations are $t=100 \rm meV$ for the
nearest neighbour hopping parameter, $B=0.45$ for next nearest
neighbour hopping, $\delta=0.1$ for the orthorhombic distortion in the
dispersion curves and $g_{x,y}=0.75t$ for the nearest neighbour
pairing potential.  The solid circles give the amplitude of the
$d_{x^2-y^2}$ component and the solid squares the extended
$s_{x^2+y^2}$ component.  The gap in $\rm meV$ is given by the left
hand side vertical axis while the phase between the two components is
given by the open circles and the scale is shown on the right hand side
vertical axis.  It varies from 0 to $\pi$.}
\label{DvsN.fig}
\end{figure}

\begin{figure}
\caption{The temperature dependence ($T$) of the gap for a case with
filling $n=0.36$ which falls in the region of finite phase $\phi$ and
complex gap in the phase diagram of Fig.~1.  The other parameters are
as given in Fig.~1.  The critical temperature is $68{\rm K}$.  The
amplitude of the $d_{x^2-y^2}$ component is given by the solid circles
and that of the $s_{x^2+y^2}$ by the solid squares.  The phase which
sets in abruptly at $T=36{\rm K}$ is shown as the open circles.
Again, the scale for the gap is on the left hand side in $\rm meV$ and
for the phase on the right hand vertical axis in degrees from (0 to
90). The inset shows the temperature dependance of the x and y
penetration depths.}
\label{DvsT.fig}
\end{figure}

\begin{figure}
\caption{Contours of the gap as a function of momentum in the first
Brillouin zone of the $\rm CuO_2$ plane.  Also shown is the Fermi line
for filling $n=0.36$.  Because of the orthorhombic distortion included
in our work through the parameter $\delta$ in the dispersion curve,
the Fermi surface does not have tetragonal symmetry and is open in the
horizontal direction and crosses the gap contours.  Frame (a) is for
$|\Delta(T=30{\rm K})|$, frame (b) for $\Delta(T=40{\rm K})$, frame
(c) for $|\Delta(T=1{\rm K})|$ and frame (d) for $|\Delta(T=40{\rm
K})|$. Note that the value of the basis functions at $(0,\pi)$ is two times the
amplitude in Fig.~2.}
\label{DvsK.fig}
\end{figure}

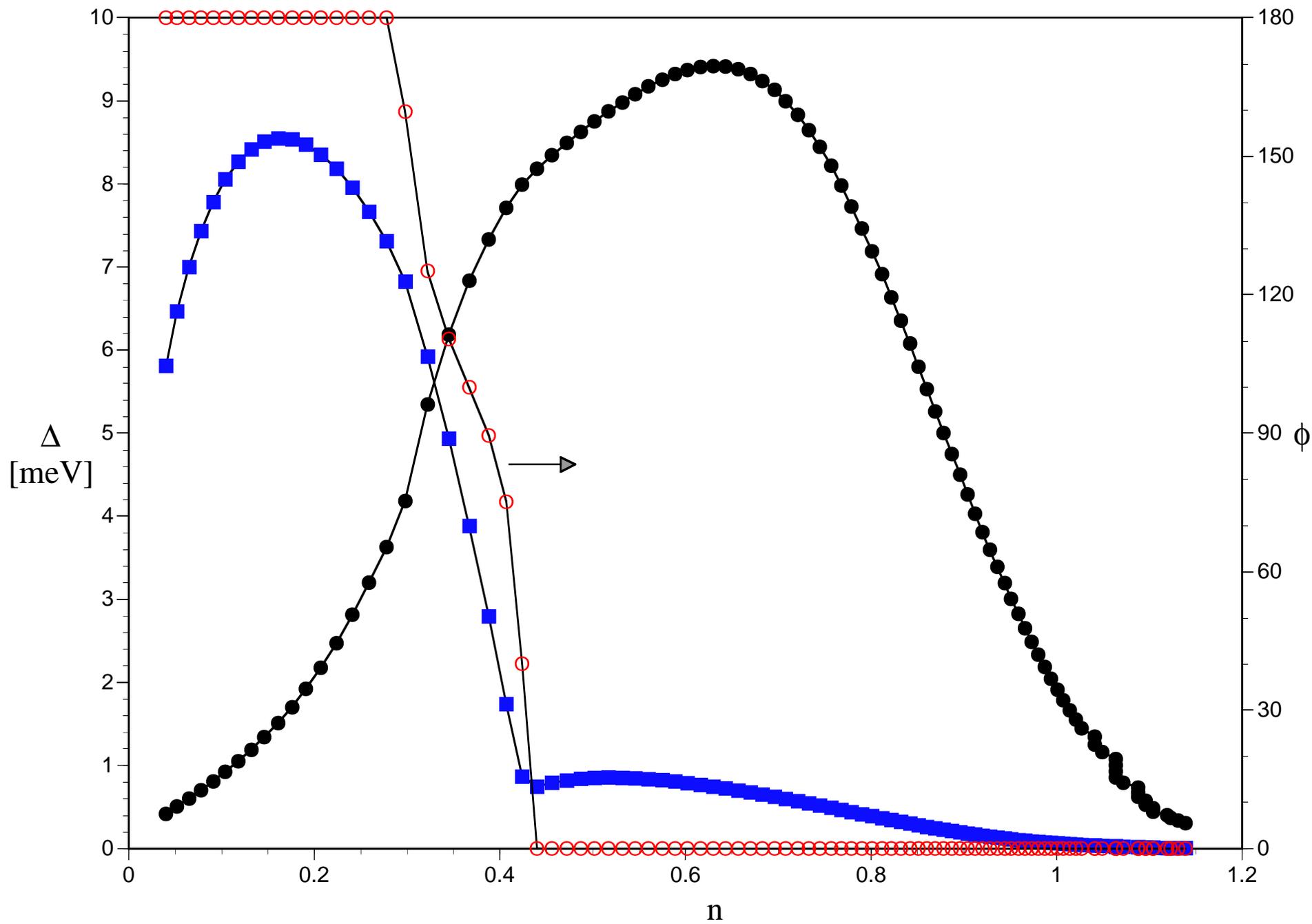
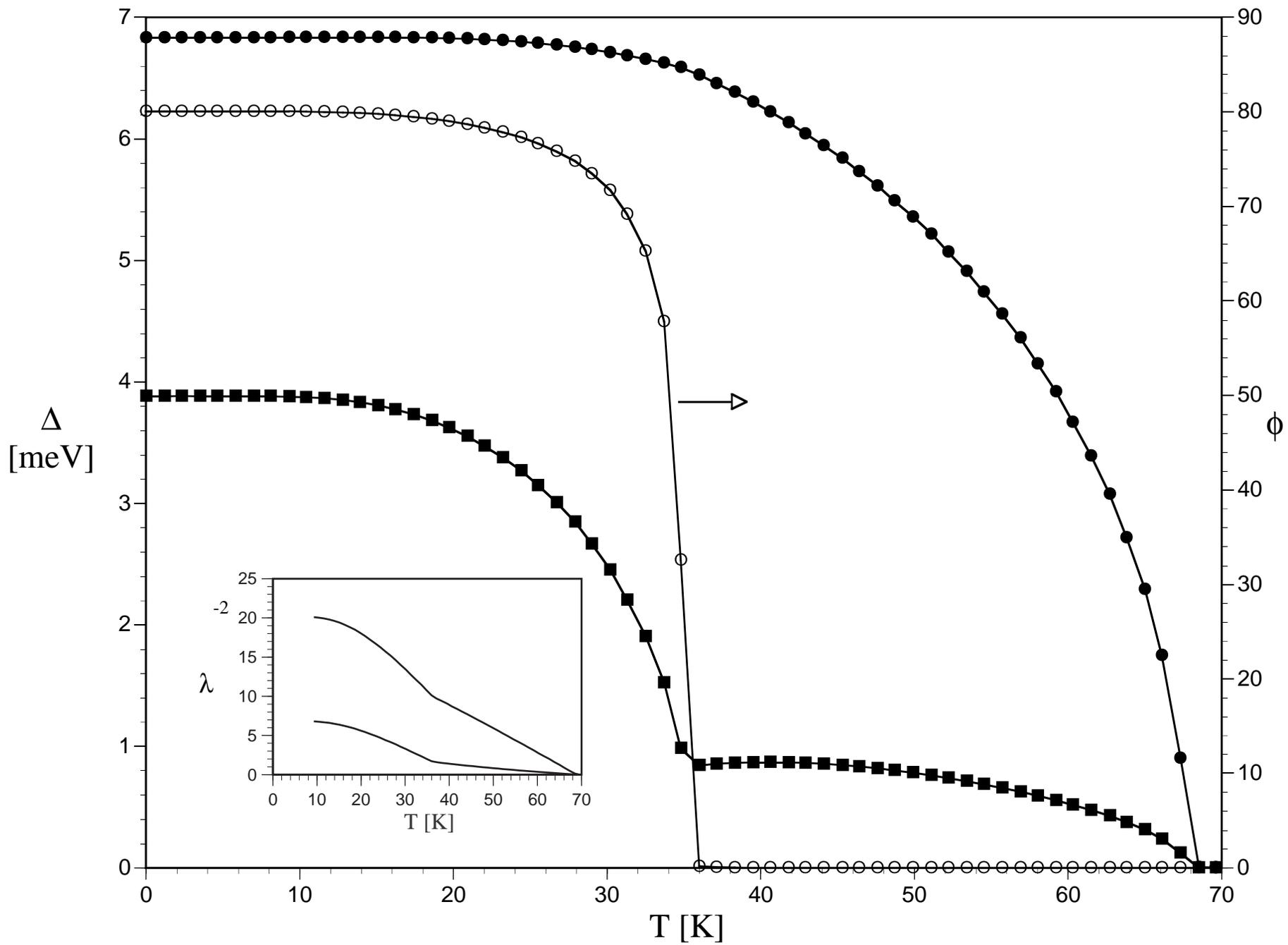
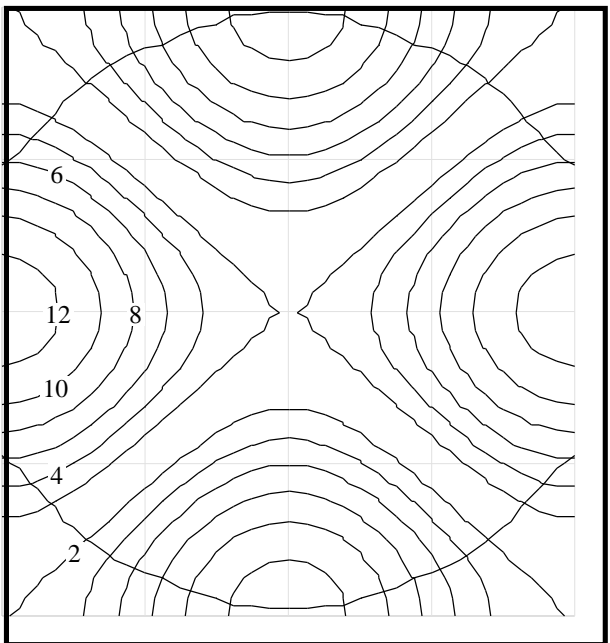
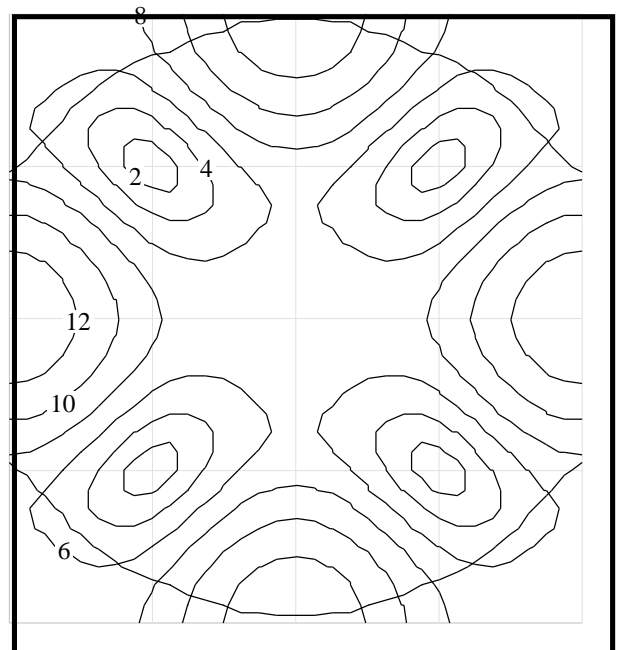
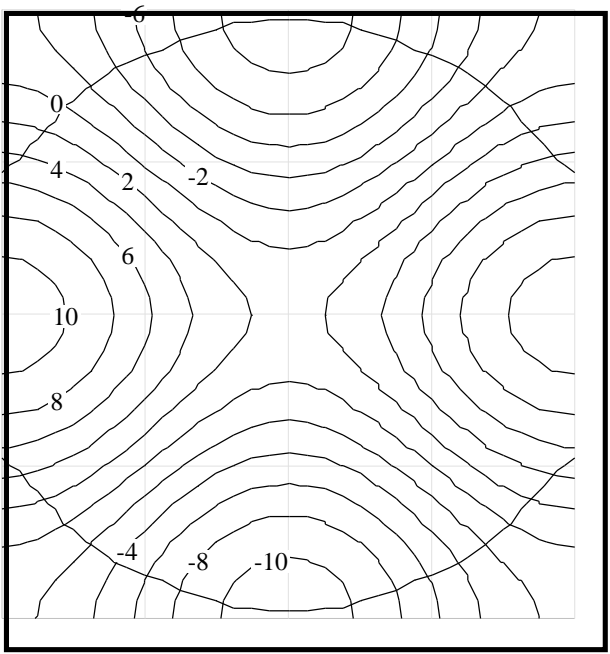
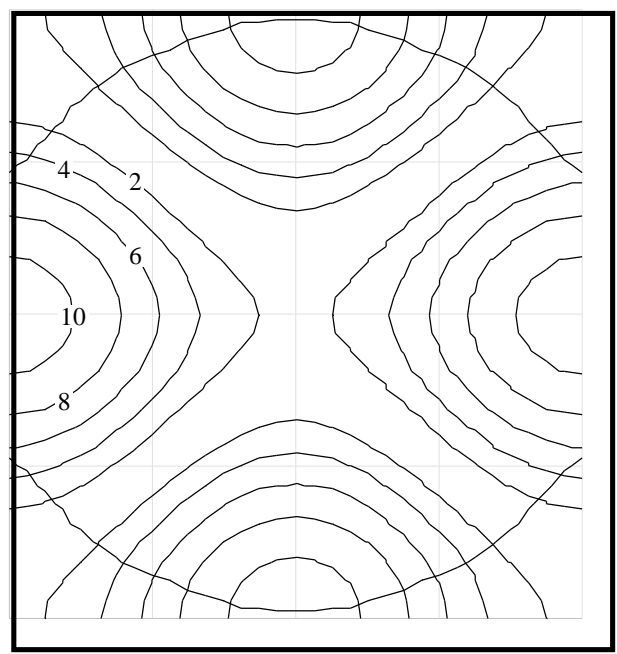
\end{document}